\pdfoutput=1
\documentclass[sigconf]{acmart}

\AtBeginDocument{%
  \providecommand\BibTeX{{%
    \normalfont B\kern-0.5em{\scshape i\kern-0.25em b}\kern-0.8em\TeX}}}

\usepackage{microtype}
\usepackage{graphicx} 
\usepackage{makecell}
\usepackage{amsmath}

\usepackage{subcaption}
\usepackage{multirow}
\begin{document}

\title{DebiasRec: Bias-aware User Modeling and Click Prediction \\ for Personalized News Recommendation}


\author{Jingwei Yi}
\affiliation{%
  \institution{University of Science and Technology of China}
}
\email{yjw1029@mail.ustc.edu.cn}

\author{Fangzhao Wu}
\affiliation{%
  \institution{Microsoft}
}
\email{wufangzhao@gmail.com}

\author{Chuhan Wu}
\affiliation{%
  \institution{Tsinghua University}
}
\email{wuchuhan15@gmail.com}

\author{Qifei Li}
\affiliation{%
  \institution{Beihang University}
}
\email{wxliqifei@sina.com}

\author{Guangzhong Sun}
\affiliation{%
  \institution{University of Science and Technology of China}
}
\email{gzsun@ustc.edu.cn}

\author{Xing Xie}
\affiliation{%
  \institution{Microsoft}
}
\email{xingx@microsoft.com}

% %%
% %% By default, the full list of authors will be used in the page
% %% headers. Often, this list is too long, and will overlap
% %% other information printed in the page headers. This command allows
% %% the author to define a more concise list
% %% of authors' names for this purpose.


%\begin{CCSXML}
%<ccs2012>
%<concept>
%<concept_id>10002951.10003260.10003261.10003271</concept_id>
%<concept_desc>Information systems~Personalization</concept_desc>
%<concept_significance>500</concept_significance>
%</concept>
%</ccs2012>
%\end{CCSXML}
%\ccsdesc[500]{Information systems~Personalization}

\keywords{News Recommendation, User Modeling, Bias}


\begin{abstract}

News recommendation is critical for personalized news access.
Existing news recommendation methods usually infer users' personal interest based on their historical clicked news, and train the news recommendation models by predicting future news clicks.
A core assumption behind these methods is that news click behaviors can indicate user interest.
However, in practical scenarios, beyond the relevance between user interest and news content, the news click behaviors may also be affected by other factors, such as the bias of news presentation in the online platform.
For example, news with higher positions and larger sizes are usually more likely to be clicked.
The bias of clicked news may bring noises to user interest modeling and model training, which may hurt the performance of the news recommendation model.

In this paper, we propose a bias-aware personalized news recommendation method named DebiasRec, which can handle the bias information for more accurate user interest inference and model training.
The core of our method includes a bias representation module, a bias-aware user modeling module, and a bias-aware click prediction module.
The bias representation module is used to model different kinds of news bias and their interactions to capture their joint effect on click behaviors.
The bias-aware user modeling module aims to infer users' debiased interest from the clicked news articles by using their bias information to calibrate the interest model.
The bias-aware click prediction module is used to train a debiased news recommendation model from the biased click behaviors, where the click score is decomposed into a preference score indicating user's interest in the news content and a news bias score inferred from its different bias features.
Experiments on two real-world datasets show that our method can effectively improve the performance of news recommendation.

\end{abstract}

\maketitle

\section{Introduction}

\begin{figure}[!t]
  \centering
  \includegraphics[width=0.4\textwidth]{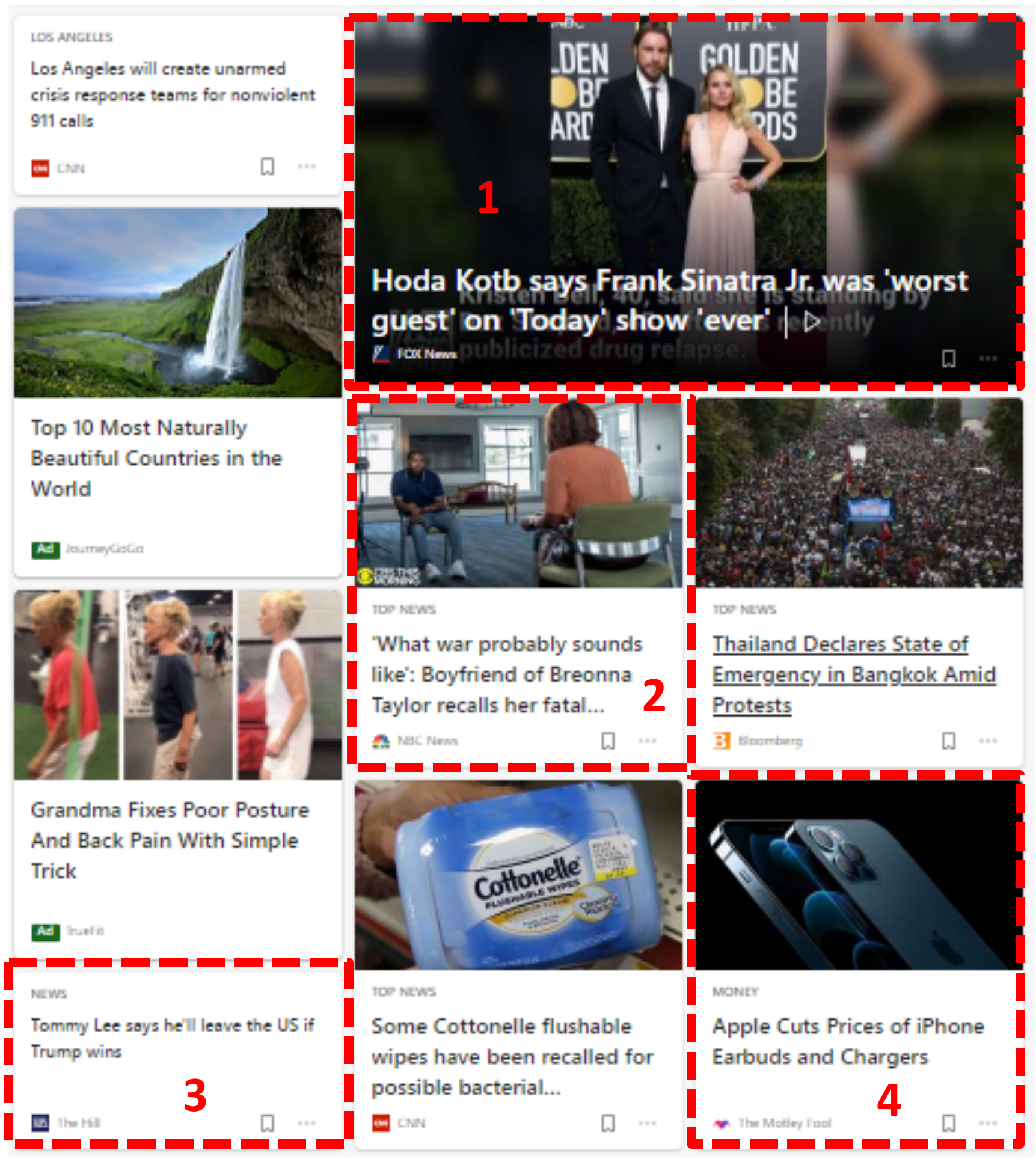}
  \caption{Illustration of different news bias (e.g., position and size) in an online news platform.}
  \label{fig:msnnews}
\end{figure}

% 新闻推荐的重要性
With the development of the World Wide Web, people's reading habits have shifted from paper-based reading to Internet reading ~\cite{liu2010personalized,wang2018dkn}.
News websites such as Google News \footnote{https://news.google.com/}, Toutiao \footnote{https://www.toutiao.com/}, and MSN News \footnote{https://www.msn.com/news} which collect and display massive news from various sources have attracted many users ~\cite{das2007google,wu-etal-2019-neural-news-recommendation}.
One vital problem for these news websites is that the large quantities of news generated every day may overwhelm users~\cite{okura2017embedding}.
Thus, it is important to utilize personalized news recommendation methods to help users find the news they are interested in and reduce their information overload~\cite{okura2017embedding,wu2019neural}.

\begin{figure}[!t]
    \centering
    \begin{subfigure}[b]{0.22\textwidth}
      \includegraphics[width=1\textwidth]{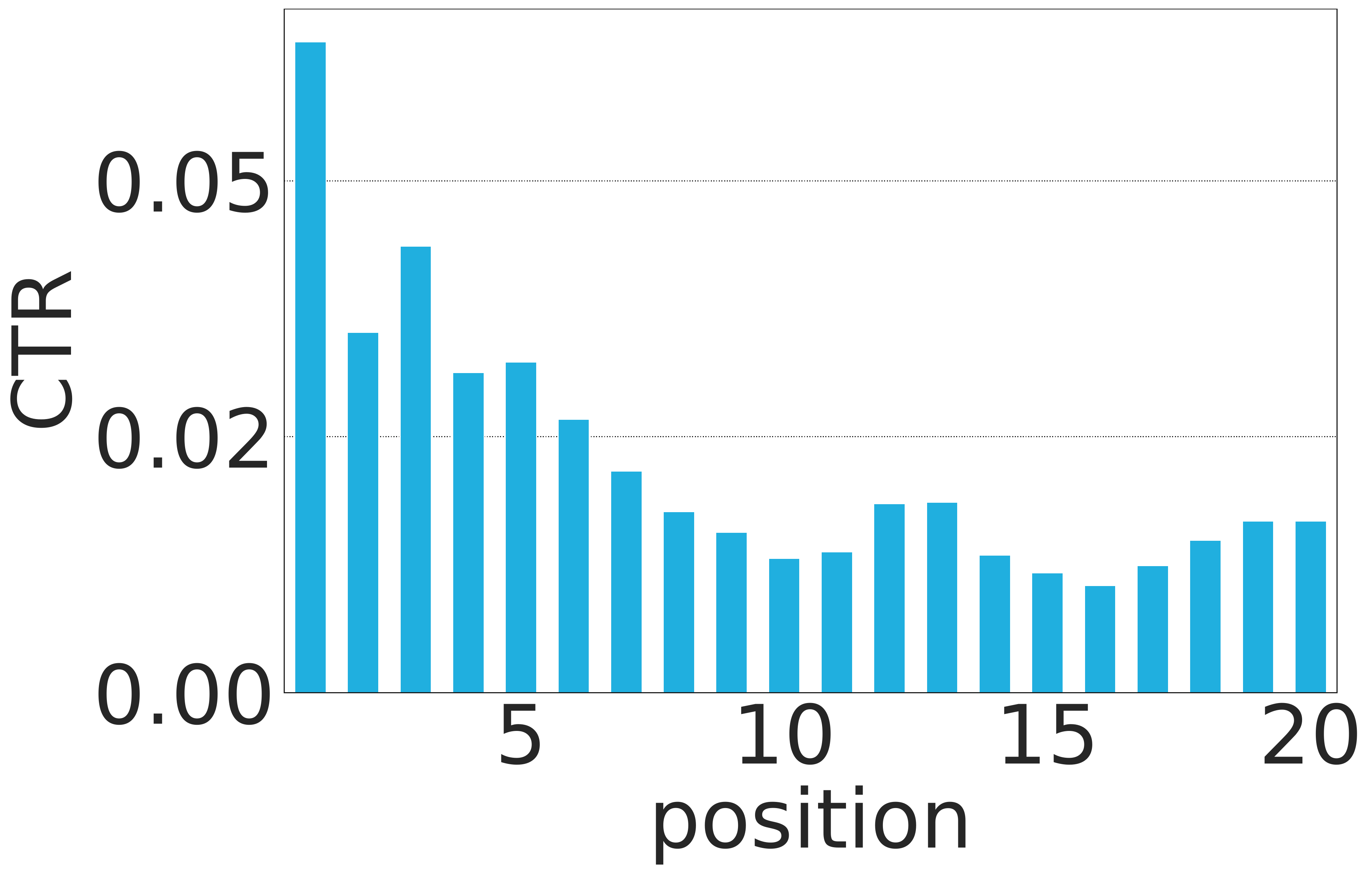}
      \caption{}
    \end{subfigure}
    \begin{subfigure}[b]{0.22\textwidth}
      \includegraphics[width=1\textwidth]{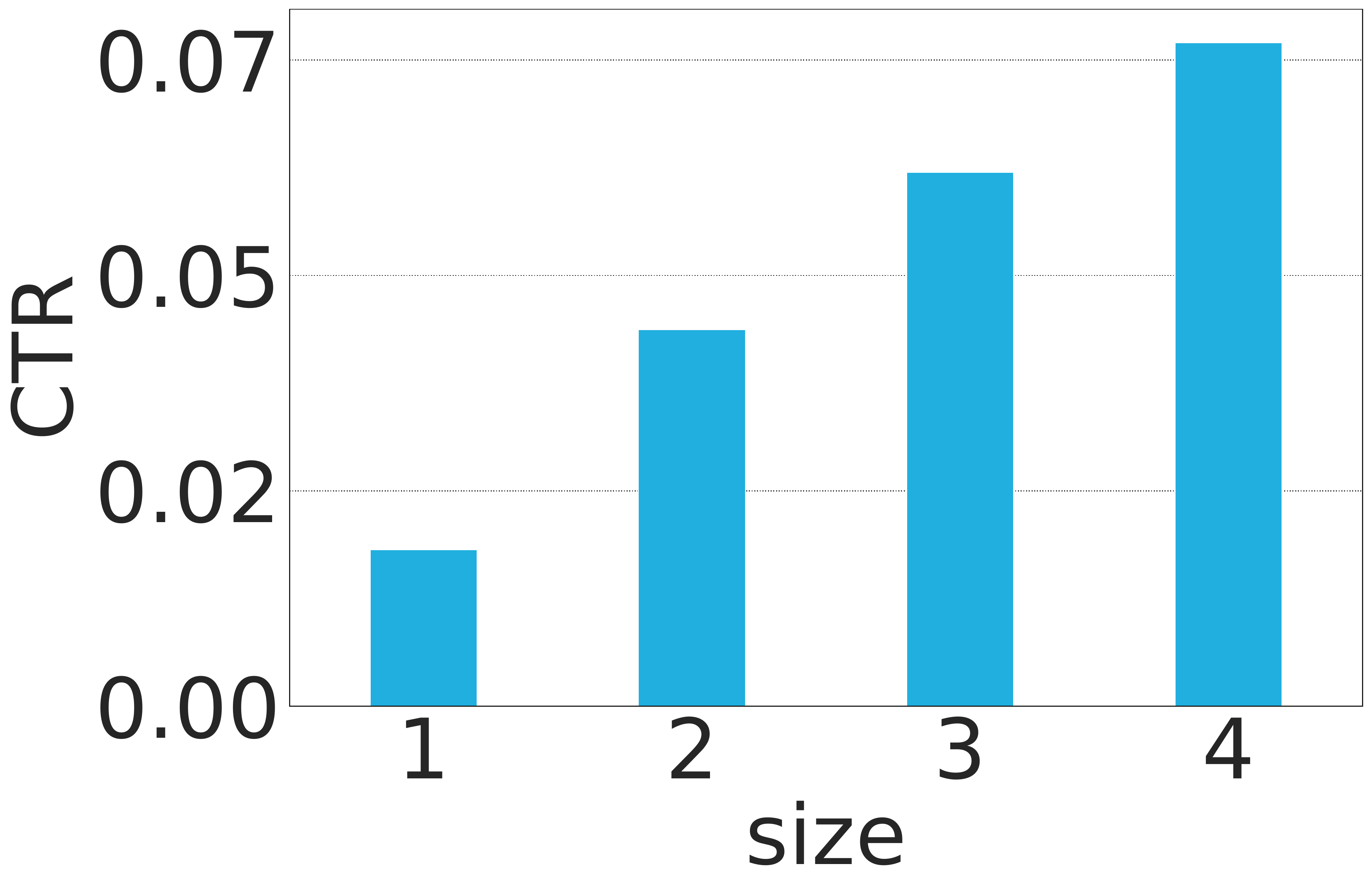}
      \caption{}
    \end{subfigure}
    \caption{Overall CTR of news at different positions and in different sizes in a commercial online news platform.}
    \label{fig:ctrpossize}
\end{figure}

News recommendation has been extensively studied in recent years~\cite{son2013location,li2014modeling,blei2003latent,wang2020fine,hu2020graph,liu2020hypernews}.
These methods usually infer user interests from the content of their clicked news articles.
For example,~\citet{wang2018dkn} used a knowledge-aware CNN to learn news representations.
A candidate-aware attention network is utilized to learn user representations from clicked news.
~\citet{wu2019neural} proposed to learn news content representations from news titles via multi-head self-attention, and they learned user interest representations from their clicked news articles via multi-head self-attention.
A core assumption behind these methods is that news click behaviors can indicate user interest.
However, users' news click behaviors may also be affected by many other factors, such as position bias~\cite{granka2008eyetracking, guan2007eye} and size bias~\cite{10.1007/978-3-319-10584-0_3}. 
For example, as shown in Figure~\ref{fig:msnnews}, the second and the forth news are displayed at different positions, while the first and the third news are displayed in different sizes.
As discussed in~\cite{richardson2007predicting}, the click-through rate (CTR) declines rapidly with lower displayed positions. 
The same phenomenon is observed in the news dataset as well.
As shown in Figure~\ref{fig:ctrpossize}, the CTR drops with lower displayed positions and increases with larger displayed sizes. 
These observations imply that biases may have a strong impact on user click behaviors because users may click the ``attractive'' news instead of the news they are really interested in~\cite{wang2020click}.
The bias in news clicks may bring noise to user interest modeling and model training, which may hurt the model performance. 
Therefore, it is critical to reduce the influence of biases in the click data for better targeting user interests.
However, existing news recommendation methods do not consider the bias effect in user modeling and model training, which may be suboptimal.

In this paper, we propose a bias-aware personalized news recommendation method named DebiasRec\footnote{Our source codes will be soon released.}, which can handle the bias information in click behaviors for both user interest inference and recommendation model training. 
Our DebiasRec method contains three core modules, i.e., the bias representation model (BRM), the bias-aware user modeling (BAUM) and the bias-aware click prediction (BACP).
The bias representation module BRM is used to learn bias representations by incorporating different kinds of bias features and their interactions to model their joint effect on click behaviors.
The bias-aware user modeling module BAUM is used to infer unbiased user interest from the biased historical news clicked behaviors.
In this module, we propose a bias-aware attention network to select informative clicked news for user interest modeling, which can distinguish whether the user is actually interested in the news content or just attracted by the presentation bias.
In the bias-aware click prediction module BACP, we propose to decompose the click prediction score into a bias score and a preference score. 
The bias score is inferred from the bias features of candidate news to model their effect on click. 
The preference score is used to measure user's personalized interest in the content of candidate news.
We only use the preference scores for candidate news ranking in online recommendation.
Extensive experiments on two real-world datasets show that DebiasRec can effectively improve the performance of news recommendation by handling the bias in user modeling and model training.

The main contributions of our work are summarized as follows:
\begin{itemize}
  \item We propose a bias-aware news recommendation method which can handle the bias information in click behaviors for accurate user interest inference and model training.
  \item We propose a bias representation model to jointly model different bias features and their interactions.
  \item We propose a bias-aware user modeling method to infer accurate user interest from the biased click behaviors, and a bias-aware click prediction method for training debiased news recommendation models.
  \item We conduct experiments on real-world datasets to verify the effectiveness of the proposed methods.
\end{itemize}

\section{Related Works}

\subsection{News Recommendation}
Personalized news recommendation is a core technique on online news platforms to improve users’ reading experience~\cite{okura2017embedding}. 
Accurate user interest modeling and click prediction are critical for news recommendation~\cite{wu2020mind}.
Existing news recommendation methods usually model user interests from their clicked news and predict click scores based on the relevance between user and candidate news representations~\cite{okura2017embedding,wu2019npa,zhu2019dan,an2019neural,wuuser,hu2020graph}.
For example, ~\citet{wang2018dkn} proposed to model user interests from clicked news based on their relevance to candidate news, and predict click scores based on candidate news and  user embeddings via a feed-forward neural network.
~\citet{wu2019neural} proposed to use an attention network to select clicked news for learning user interest representations, and they used the inner product between candidate news and user representations to predict click scores.
~\citet{wu-etal-2019-neural-news-recommendation} proposed to use multi-head self-attention networks to model the relatedness between clicked news, and they used an attention network to select informative news clicks for learning user representations.
The click scores are also computed by the inner product between user and news embeddings.
These news recommendation methods assume that the click behaviors of users can indicate user interests~\cite{wang2020fine}.
However, users may click news not only due to the interest in the news title, but also the attraction of several biases such as large sizes and top positions~\cite{ craswell2008experimental,chen2012beyond,wang2020click}.
The bias information encoded in news click behaviors will propagate to user interest modeling and model training, which is harmful for modeling performance and user experience.
Different from these methods, our approach can handle and alleviate the bias information for both user interest inference and recommendation model training.

\subsection{Bias-aware CTR Prediction}

Click logs usually provide implicit but abundant user feedback, thereby they are widely used for modeling user interests and training recommendation models~\cite{wang2019kgat, wu2020sse, zhao2020effective}.
% However, click logs are often biased and cannot fully reflect user preferences, because there are many interest-independent factors such as presentation order and size that have strong impacts on user click behaviors~\cite{pan2004determinants, guan2007eye, granka2008eyetracking}.
However, click logs are often biased and cannot fully reflect user preferences ~\cite{pan2004determinants, guan2007eye, granka2008eyetracking}.
% Recommendation models directly trained on biased click data may inherit the bias information and generate biased recommendation results, e.g., recommending items that are used to be displayed in larger sizes and higher positions, which is not beneficial for user experience.
Thus, it is important to eliminate the bias effect in click data.
Several early methods for click data debiasing are based on probabilistic click models~\cite{chapelle2009dynamic,dupret2008user}, which are initially proposed to solve the click bias problem in web search scenario for calculating the real relevance between query and document in web search.
For example, ~\citet{craswell2008experimental} propose a cascade model, which is one of the earliest click modeling methods.
They assumed that users examine the documents sequentially and stop when the first click happens, and they used a Bayesian Network to calculate the unbiased relevance score.
However, the cascade model is not suitable for modeling the effect of other kinds of bias like vertical bias, where items are not displayed in a sequential way.
Based on the cascade hypothesis, \citet{chen2012beyond} and~\citet{ wang2013incorporating} further proposed click models which can handle both position bias and vertical bias.
However, these methods assume that the same pair of query and user appears multiple times in the training set, which is not suitable for the recommendation scenario because users usually do not click the same item repeatedly.

Several methods explored to treat the bias as a counterfactual effect for learning unbiased learning-to-rank systems~\cite{wang2016learning,joachims2017unbiased}. 
They usually rely on Inverse Propensity Weighting (IPW) to estimate the propensity weight with randomization.
However, displaying news randomly will affect user reading experience and it is not suitable for these scenarios lack of random dataset.
To resolve the problem, ~\citet{wang2018position} proposed to apply a regression-based EM approach to estimate the propensity in unbiased learning-to-rank frameworks.
However, these methods do not consider the bias effect in user interest modeling.

Several other methods employ position models to reduce the effect of position bias in recommendation~\cite{craswell2008experimental}.
Their basic idea is incorporating position information of items in the model training stages to model its effect, and deactivate it in the test stage.
For example, ~\citet{ling2017model} proposed to use the item position as the model input in the training stage, and replace it with a default position number in the inference stage.
However, the ranking results may be different when inputting different default positions.
~\citet{guo2019pal} proposed to model the click probability as the multiplication of the relevance score and the probability that a user observes the item.
They used a position module to compute observation probability based on positions and remove the position module in the online inference stage.
These methods only consider a single kind of bias, i.e., position bias, and they are not able to eliminate the joint effect of multiple kinds of bias.
Different from them, our approach can model the effect of multiple kinds of bias and meanwhile consider their interaction, which can help better eliminate bias information in user interest modeling and model training.

% Various methods~\cite{ling2017model,chapelle2009dynamic} have been proposed to address the bias problem.
% Guo et al.~\cite{guo2019pal} proposed model the click probability on two factors: a) the probability that the item is seen by the user and b) the probability that the user clicks on the item, given that the item has been seen by the user. 
% Chapelle et al.~\cite{chapelle2009dynamic} propose users read items sequentially and stop reading when they feel satisfied with the content. 
% However, these methods do not consider the bias impact in user historical behaviors and do not model the interaction between different biases.

\section{Problem Formulation}
\label{sec:problem}
% Given a user $u$, $u$ has clicked $M$ news which is denoted as $\{n_1, n_2 ... n_M\}$.
% We denote the candidate news set for user $u$ as $\{n^c_1, n^c_2 ... n^c_N\}$, and the click labels of the candidate news are denoted as $\{y_1, y_2 ... y_N\}$.
% For every click record, we have its news content information, position information and size information in our click logs, which can be denoted as $n = (t, p, s)$.
% It is noted that the click labels are biased which cannot fully represent user interest.
% In the offline training stage, the news recommendation models use the click labels as targets and predict click scores of candidate news which are denoted as $\{s_1, s_2 ... s_N\}$. 
% The bias-aware news recommendation models need to predict preference scores which are merely related to user interest towards news content.
% The preference scores are denoted as $\{s_1^p, s_2^p ... s_N^p\}$.
% In the online inference stage, recommendation systems rank news by the preference scores and display news at different positions and in different sizes.
In this section, we present a formal definition of the problem studied in this paper.
The task of our news recommendation model is predicting whether a user $u$ will click on a candidate news $n_c$ based on the content information of candidate news and the user interests inferred from clicked news.
Each news is represented by its title texts.
In addition, we assume the user $u$ has $M$ clicked news, which is denoted as $[n_1, n_2, ..., n_M]$ (sorted by the their click time).
Each historical click $n_i$ is associated with a position $p_i$ and a size $s_i$ for display.
The recommendation model needs to compute a click score $s_c$ for each pair of user and candidate news $(u, n_c)$, which is further used to rank candidate news according to user interests.
Since user click behaviors are heavily affected by many bias information such as the position and size of news, both the user interests inferred from click behaviors and the click score for news ranking may also be biased.
Thus, the goal of our approach is eliminate the bias information in user interest modeling and click prediction to help better target user interests for news recommendation.

\section{Methodology}

\begin{figure*}[!t]
  \centering
  \includegraphics[width=0.95\textwidth]{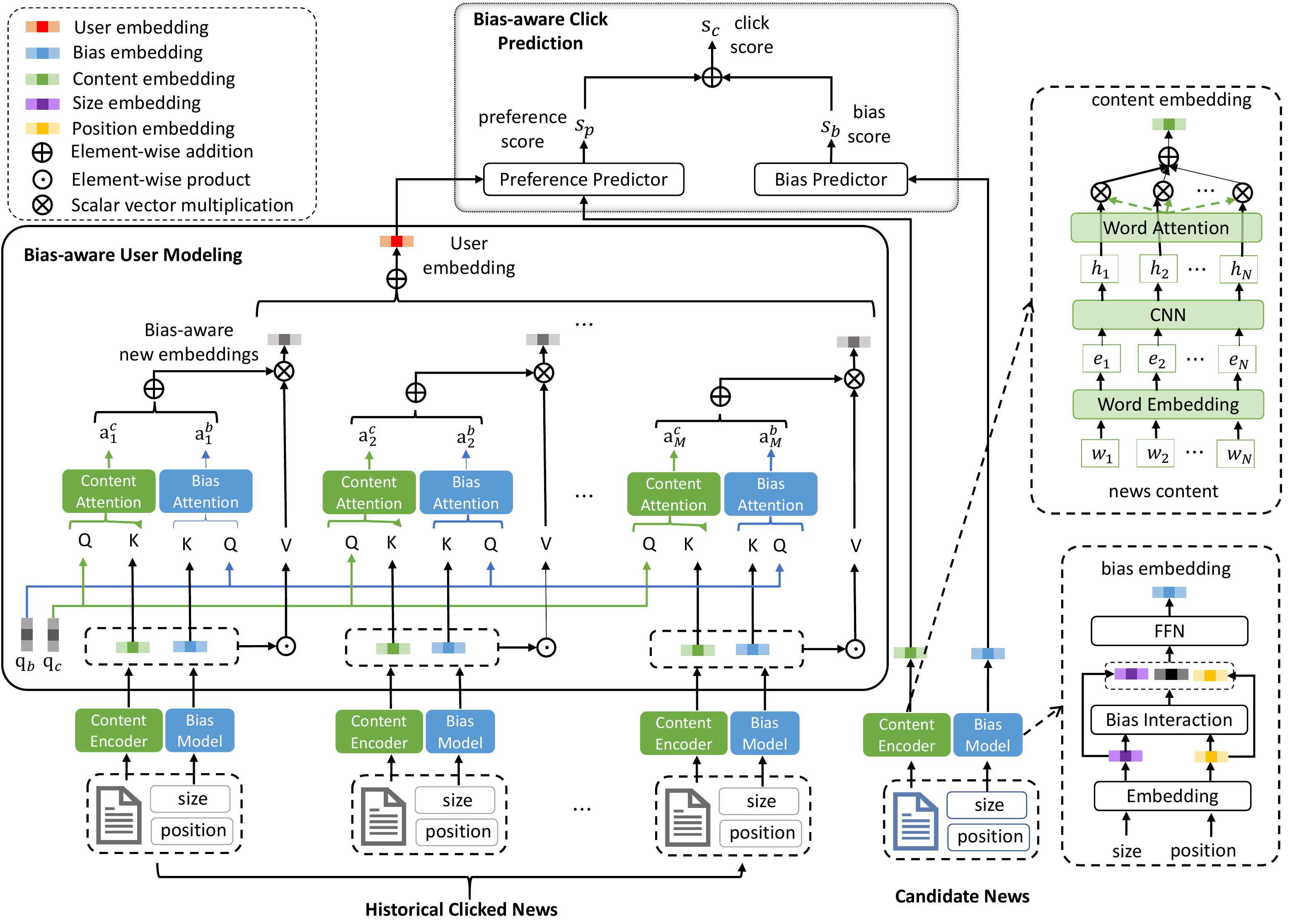}
  \caption{ The framework of our \textit{DebiasRec} method.}
  \label{fig:baum}
\end{figure*}

In this section, we introduce our bias-aware user modeling and click prediction method for news recommendation (DebiasRec).
It considers multiple kinds of biases in news clicks and can reduce their effects on user interest modeling and click prediction for better targeting user interests.
Figure~\ref{fig:baum} shows the overall framework of our DebiasRec model. 
Our DebiasRec approach contains four modules, i.e. content encoder, bias representation model (BRM), bias-aware user modeling (BAUM), and bias-aware click predictor (BACP).
Their details are introduced as follows.

\subsection{Content Encoder}
\label{sec:content-encoder}
In this section, we introduce the content encoder module in DebiasRec, which is based on the title encoder in NAML~\cite{wu2019neural}.
The content encoder module is used to learn news content representations, which contains three layers.
The first layer is an embedding layer, which converts the words within a news title into a sequence of low-dimensional word embeddings.
The second layer is a convolutional neural network (CNN), which captures the local contexts of each word to learn their contextual representations.
For the $i_{th}$ word in the news title, its contextual word representation learned by this CNN layer is denoted as $\textbf{h}_i^w$.
The third layer is a word-level attention network, which is used to select important words within the same news title for learning informative news representations.
We denote the attention weight of the $i_{th}$ word as $\alpha_i^w$, which is computed as follows:
\begin{equation}
  \begin{aligned}
  &a_i^w = \textbf{q}_w^T \tanh(\textbf{V}_w\times \textbf{h}_i^w + \textbf{v}_w), \\
  &\alpha_i^w = \frac{\exp(a_i^w)}{\sum_{j=1}^{L}\exp(a_j^w)},
  \end{aligned}
\end{equation}
where $q_w$ is a learnable query vector in the word-level attention network, $V_w$ and $v_w$ are projection parameters, and $L$ is the length of the news title.
The final representation $\textbf{c}$ of a news title is the weighted summation of its contextual word representations, i.e., $\textbf{c} = \sum_{i=1}^{L} \alpha_i^w\times \textbf{h}_i^w$.

\subsection{Bias Representation Model}
\label{biasmodel}
Next, we introduce the bias representation model (BRM) in our approach, which aims to learn accurate bias representations from multiple kinds of bias information and meanwhile consider their interactions.
As shown in Figure~\ref{fig:ctrpossize}, the position and the size have strong impacts on user click behaviors.
Therefore, in our bias representation model, we model the impacts of both biases.
For the position $p$ of each news, we use a position embedding layer to convert it into a low-dimensional position embedding $\textbf{e}_p$.
For the size of news $s$, we use another size embedding layer to convert it into its size embedding  $\textbf{e}_s$.
A simple way to generate bias representations is to utilize a linear perceptron layer to transform the concatenate of $\textbf{e}_p$ and $\textbf{e}_s$ into a unified bias representation $\textbf{b}$. 
The bias representation $\textbf{b}$ can be computed as follows: 
\begin{equation}
\label{linear_bias_model}
    \textbf{b} = \textbf{W}_b \times [\textbf{e}_p || \textbf{e}_s] + \textbf{b}_b,
\end{equation}
where $||$ is concatenate operation, $\textbf{W}_b$ and $\textbf{b}_b$ are projection parameters.
However, the position bias and the size bias are not independent of each other (verified by the Chi-square test result in Section~\ref{sec:biasinter}).
For example, large news is likely to be displayed at special and conspicuous positions, i.e., in the middle or top of the screen.
Researchers have found that the add operation of the perceptron layer used in Eq.~\ref{linear_bias_model} is  weak in capturing the interactions between variables~\cite{qu2016product, menon2011response,  ta2015factorization}.
The bias effects may not be effectively modeled if their interactions are not fully considered.
Inspired by~\citet{qu2016product}, we use the element-wise product to help capture the dependencies between different biases.
In our method, the unified bias representation  $\textbf{b}$ is computed as follows:
\begin{equation}
\begin{aligned}
    &\textbf{e}_{i} = \textbf{e}_p \odot \textbf{e}_s, \\
    &\textbf{b} = \textbf{W}_b \times [\textbf{e}_p || \textbf{e}_s || \textbf{e}_{i}] + \textbf{b}_b,
\end{aligned}
\end{equation}
where $\odot$ stands for the element-wise product operation.
In this way, the bias representation encodes not only the information of each kind of bias, but also the interactions between different kinds of biases, which may be helpful for effective debiasing.

\subsection{Bias-aware User Modeling}

The bias-aware user modeling (BAUM) module is used to learn debiased user interest representations from  click behaviors by distilling the real user preferences from them with the help of bias information.
Users' click behaviors do not necessarily indicate user interests because they are heavily influenced by many bias information.
Users may click news due to the attraction of news displayed at top positions or in large size rather than their personal interests.
For example, in Figure~\ref{fig:ctrpossize}, a user may  be ``attracted'' to click the first news, which is the most conspicuous one in the webpage.
It is difficult to identify whether the user is actually interested in the content of this news.
However, if a user clicks the third news ``Tommy Li says he will leave the US if trump wins'' which is displayed with small size and low position, we can infer that the user is probably interested in this news.
Thus, selecting clicked news that can better indicate user interests is important for reducing the bias information in user modeling and calibrating the interest model.

Motivated by these observations, we propose a bias-aware attention network to debias the click behaviors and learn debiased user interest representations.
For each clicked news  $n_i$, we first compute its content representation $\mathbf{c}_i$ using the content encoder and compute its bias representation $\textbf{b}_i$ according to its position $p_i$ and size $s_i$ using the bias representation model.
The bias-aware attention network takes both the content and bias representations of clicked news as input to calculate bias attention scores and content attention scores.
The bias attention mechanism aims to estimate the expected influence of bias factors on click probabilities and reduce bias effects on user modeling.
We denote the bias attention score of the $i$-th news as $a_i^b$, which is computed as follows:
\begin{equation}
a_i^b = \textbf{q}_b^T \tanh(\textbf{V}_b\times \textbf{b}_i + \textbf{v}_b),
\end{equation}
where $\textbf{q}_b$ is a learnable query vector in the bias attention network, and $\textbf{V}_b$ and $\textbf{v}_b$ are projection parameters.

The content attention mechanism aims to evaluate the informativeness of the content of clicked news for user interest modeling.
According to previous studies, different news usually contributes differently toward modeling user interests~\cite{zhu2019dan,wu2019neural,wu-etal-2019-neural-news-recommendation}.
For example, the title `Pulse Extra' is much less informative for inferring user interest than another title `Spectacular photos from space'.
Therefore, we use a content attention network to select informative click behaviors.
It computes the content attention scores of clicked news based on the their content representations. 
We denote the content attention score of the $i$-th clicked news as $a_i^c$, which is computed as follows:
\begin{equation}
a_i^c = \textbf{q}_c^T \tanh(\textbf{V}_c\times \textbf{c}_i + \textbf{v}_c),
\end{equation}
where $\textbf{q}_c$ is the query vector of content attention network, and $\textbf{V}_c$ and $\textbf{v}_c$ are parameters.

Then, the content attention scores and bias attention scores are fused into unified attention weights that indicate the relative importance of different clicked behaviors.
We denote the unified attention weight of the $i$-th click behavior as $\alpha_i^n$, the user representation as \textbf{u}, which are calculated as follows:
\begin{equation}
  \alpha_i^n = \frac{\exp(a_i^c+a_i^b)}{\sum_{j=1}^{M}\exp(a_j^c+a_j^b)},
 \end{equation}
The final user representation $\textbf{u}$ is the weighted summation of the element product between the content and bias representations of  clicked news, which is formulated as:
\begin{equation}
\textbf{u} = \sum_{i=1}^{M}{\alpha_i^n \times (\textbf{b}_i \odot \textbf{c}_i}),
\end{equation}
where $\odot$ represents the element-wise product operation.

\subsection{Bias-aware Click Prediction}

The bias-aware click prediction (BACP) module aims to help learn a debiased news recommendation model.
Besides the user interests, some interest-independent bias factors also strongly influence user click behaviors, such as the displayed position and size of news.
As shown in Figure~\ref{fig:ctrpossize}, the click-through rate decreases rapidly with the positions and increases with the sizes.
Several eye tracking studies ~\cite{granka2008eyetracking,guan2007eye,pan2004determinants} also show the same conclusion.
Therefore, click labels contain much bias information, and the directly trained with the biased click data will inherit the unwanted biases, which is harmful for accurate user interest matching.
To solve this problem, we propose a bias-aware click predictor to model the bias effects on click labels and  compute bias-free user preference scores.
In the bias-aware click predictor, we decompose the overall click score ${s_c}$ into a summation of a preference score ${s_p}$ that indicates user interests on a candidate news and a bias score ${s_b}$ that indicates the influence of bias on click probability.
The preference score ${s_p}$ is computed by the dot product between the user representation and candidate news representation, and the bias scores ${s_b}$ is predicted from the bias representation $\textbf{b}$ of candidate news, which are formulated as follows:
\begin{equation}
  \begin{aligned}
  &{s_b} = \textbf{W}_{pb} \times \textbf{b} + \textbf{b}_{pb},\\
  &{s_p} = \textbf{u}^T \textbf{c} ,\\
  &{s_c} = {s_b} + {s_p},
  \end{aligned}
\end{equation}
where $\textbf{W}_{pb}$ and $\textbf{b}_{pb}$ are projection parameters.
Note that only the preference score ${s_p}$ is used to rank candidate news in the test stage, i.e., only  historical clicked behaviors and the content information of candidate news are used for test.
In this way, DebiasRec can generate debiased click prediction scores for recommendation.

\subsection{Model Training}
Following many previous methods~\cite{huang2013learning, wu-etal-2019-neural-news-recommendation}, we apply negative sampling strategy to build training samples. 
For every positive sample (clicked news), we randomly sample $K$ negative samples (non-clicked news) from the same impression.
We denote the prediction click scores, preference scores and bias scores of these $K+1$ sample as $\{(s_c^i, s_b^i, s_p^i)|i=1,2...K+1\}$, and denote the click labels of the samples as $\{y^1, y^2, ... y^{K+1}\}$ (1 for click and 0 for non-click). 
In the training process, we first use the softmax function to normalize each click score into a click probability, which is computed as:
\begin{equation}
  p^i = \frac{exp(s_c^i)}{\sum_{k=1}^{K+1}exp(s_c^k)},
\end{equation}
To maximize the click probability of the positive sample, we use the cross-entropy loss as the loss function, which is formulated as:
\begin{equation}
  \mathcal{L} = -\sum_{i=1}^{K+1}y^i\times log(p^i).
\end{equation}
% In the online inference stage, the position and size features of candidate news are determined after generating a ranking list. 
% Therefore, bias features are unavailable.
% We only use the preference scores to recommend news.

% Motivated by \cite{huang2013learning}, we use negative sampling technology for model training.
% For each clicked news (regard as a positive sample), we will random sample $K$ news which is not clicked by the user in the same impression(regard as negative samples).
% Our model will join predict the click probability scores of these $K+1$ samples. We formulate the task as a pseudo $K+1$-way classification task. 
% We minimize the summation of the negative log-likelihood
% of all positive samples during training, which can
% be formulated as follows: 
% \begin{equation}
%   -\sum_{i=1}^{P}log\frac{exp(y_i^p)}{exp(y_i^p) + \sum_{k=1}^{K}exp(y_{i,k}^n)},
% \end{equation}
% where $P$ is the number of positive samples in training dataset. $y_i^p$ is the click score of the $i_th$ positive sample, and $y_{i,k}^n$ is the click score of the k-th negative sample in the
% same impression with the i-th positive sample.

% Since the position and size feature of candidate news is inaccessible in online inference stage, we only use preference score $\widehat{y_p}$ to evaluate model effectiveness.

\begin{table*}[h]
  \centering
  \caption{Results of different methods. *Improvements over other baselines are significant ($p<0.05$ in t-test).}
  \resizebox{0.9\textwidth}{!}{
  \begin{tabular}{c|cccc|cccc}
  \Xhline{1.5pt}
  \hline
             & \multicolumn{4}{c|}{\textit{News}}                                                                 & \multicolumn{4}{c}{\textit{Feeds}}                                                                   \\ \hline
  Methods    & AUC                                & MRR                                & nDCG@5         & nDCG@10        & AUC                                & MRR                                & nDCG@5         & nDCG@10        \\ \hline
  LibFM      & 58.26$\pm$0.06                              & 19.04$\pm$0.04                              & 18.91$\pm$0.04          & 23.48$\pm$0.04          & 60.82$\pm$0.02                              & 28.57$\pm$0.03                              & 30.20$\pm$0.04          & 38.18$\pm$0.02          \\
  DSSM       & 60.95$\pm$0.15                              & 20.25$\pm$0.14                              & 20.35$\pm$0.12          & 25.06$\pm$0.12          & 62.30$\pm$0.33                              & 29.71$\pm$0.35                              & 31.59$\pm$0.41          & 39.56$\pm$0.36          \\
  DeepFM     & 58.32$\pm$0.35                              & 18.54$\pm$0.26                              & 18.40$\pm$0.25          & 23.09$\pm$0.25          & 62.56$\pm$0.16                              & 30.06$\pm$0.11                              & 32.00$\pm$0.12          & 39.87$\pm$0.14          \\
  Wide\&Deep & 60.05$\pm$0.17                              & 19.59$\pm$0.21                              & 19.61$\pm$0.24          & 24.33$\pm$0.23          & 62.82$\pm$0.05                              & 30.07$\pm$0.08                              & 32.05$\pm$0.11          & 39.97$\pm$0.09          \\
  DFM        & 60.22$\pm$0.08                              & 19.78$\pm$0.15                              & 19.82$\pm$0.14          & 24.50$\pm$0.12          & 62.71$\pm$0.16                              & 30.17$\pm$0.11                              & 32.15$\pm$0.15          & 39.99$\pm$0.11          \\
  DKN        & 61.14$\pm$0.15                              & 19.99$\pm$0.34                              & 20.02$\pm$0.29          & 24.87$\pm$0.28          & 63.37$\pm$0.15                              & 30.44$\pm$0.16                              & 32.53$\pm$0.20          & 40.42$\pm$0.17          \\
  NAML       & 62.65$\pm$0.20                              & 21.55$\pm$0.13                              & 21.94$\pm$0.15          & 26.76$\pm$0.17          & 64.10$\pm$0.12                              & 31.20$\pm$0.07                              & 33.46$\pm$0.06          & 41.27$\pm$0.06          \\
  NRMS       & 62.76$\pm$0.18                              & 21.56$\pm$0.16                              & 21.98$\pm$0.17          & 26.84$\pm$0.19          & 63.91$\pm$0.36                              & 31.05$\pm$0.30                              & 33.24$\pm$0.36          & 41.09$\pm$0.34          \\ \hline
  REM        & 62.88$\pm$0.21                              & 21.66$\pm$0.14                              & 22.24$\pm$0.17          & 26.89$\pm$0.22          & \multicolumn{1}{l}{64.34$\pm$0.15}          & \multicolumn{1}{l}{31.37$\pm$0.17}          & 33.61$\pm$0.13          & 41.43$\pm$0.16          \\
  PAL        & \multicolumn{1}{l}{62.96$\pm$0.36}          & \multicolumn{1}{l}{21.74$\pm$0.19}          & 22.13$\pm$0.24          & 26.97$\pm$0.24          & \multicolumn{1}{l}{64.45$\pm$0.09}          & \multicolumn{1}{l}{31.45$\pm$0.06}          & 33.77$\pm$0.06          & 41.55$\pm$0.07          \\ \hline
  DebiasRec*      & \multicolumn{1}{l}{\textbf{63.70}$\pm$0.15} & \multicolumn{1}{l}{\textbf{22.33}$\pm$0.17} & \textbf{22.81}$\pm$0.17 & \textbf{27.67}$\pm$0.15 & \multicolumn{1}{l}{\textbf{65.34}$\pm$0.07} & \multicolumn{1}{l}{\textbf{32.31}$\pm$0.10} & \textbf{34.83}$\pm$0.12 & \textbf{42.57}$\pm$0.10 \\ \hline
   \Xhline{1.5pt}
  \end{tabular}
  }
  \label{tab:baslines}
\end{table*}

\begin{table}[!t]
\centering
\caption{Detailed statistics of \textit{News} and \textit{Feeds} datasets.}
\resizebox{0.48\textwidth}{!}{
\begin{tabular}{cccccc}
\Xhline{1.5pt}
                & \# News & \# Users & \# Impressions & \# Click & \# Sizes \\ \hline
\textit{News}  & 97,646  & 310,469  & 58,000,000     & 917,839  & 4             \\
\textit{Feeds} & 643,177 & 10,000   & 320,925        & 970,846  & 4             \\ 
\Xhline{1.5pt}
\end{tabular}
}
\label{tab:statistic}
\end{table}

\section{Experiments}
In this section, we demonstrate the effectiveness of our approach by answering the following four research questions:
\begin{itemize}
    \item \textbf{RQ1:} How does DebiasRec perform compared with baseline methods?
    \item \textbf{RQ2:} Are BAUM and BACP both useful?
    \item \textbf{RQ3:} Can BRM model different biases and their interactions for accurate news bias modeling?
    \item \textbf{RQ4:} Can DebiasRec provide reasonable recommendation results?
    
\end{itemize}

\subsection{Datasets and Experimental Settings}

Since there is no off-the-shelf news recommendation dataset with position bias and presentation bias information, we build two real-world datasets from the user logs of two news platforms, named \textit{News} and \textit{Feeds}.
\textit{News} is collected on Microsoft News website from October 13th, 2019 to November 13th, 2019.
\textit{Feeds} is collected on Microsoft News App from August 1st, 2020 to September 1st, 2020.
The detailed dataset statistics are summarized in Table~\ref{tab:statistic}.
For both datasets, the impressions in the first three weeks are used for model training, and the rest impressions in the last week are used for test.
Additionally, we randomly sample 20\% impressions in the training set for validation. 
Note that the bias features of candidate news are not used for click prediction in the test stage, while the bias features of user historical clicked news are used  in both training and test stages to debias user click behaviors.
Following many previous news recommendation works ~\cite{wu2019neural, wu2020mind}, we use AUC, MRR, nDCG@5 and nDCG@10 as the evaluation metrics.

In our experiments, we apply pretrained Golve embedding~\cite{pennington2014glove} to initialize the word embedding matrix. 
The dimension of word embedding vectors is 300.
The dimension of position and size embedding vectors is 200.
The number of filters in CNN layer is 400, and the window size is 3. 
We apply dropout to each layer to mitigate overfitting.
The dropout rate is 0.2.
Adam~\cite{kingma2014adam} is used for model optimization.
The batch size is set to 32.
The number of negative samples associated with each positive sample is 4.
The max position is set to 400 in both datasets.
All hyper-parameters are selected according to results on the validation set.
We repeat each experiment 5 times independently, and we report the average results with standard deviations.

\begin{figure*}[!t]
    \centering
    \begin{subfigure}[b]{0.49\textwidth}
      \includegraphics[width=0.9\textwidth]{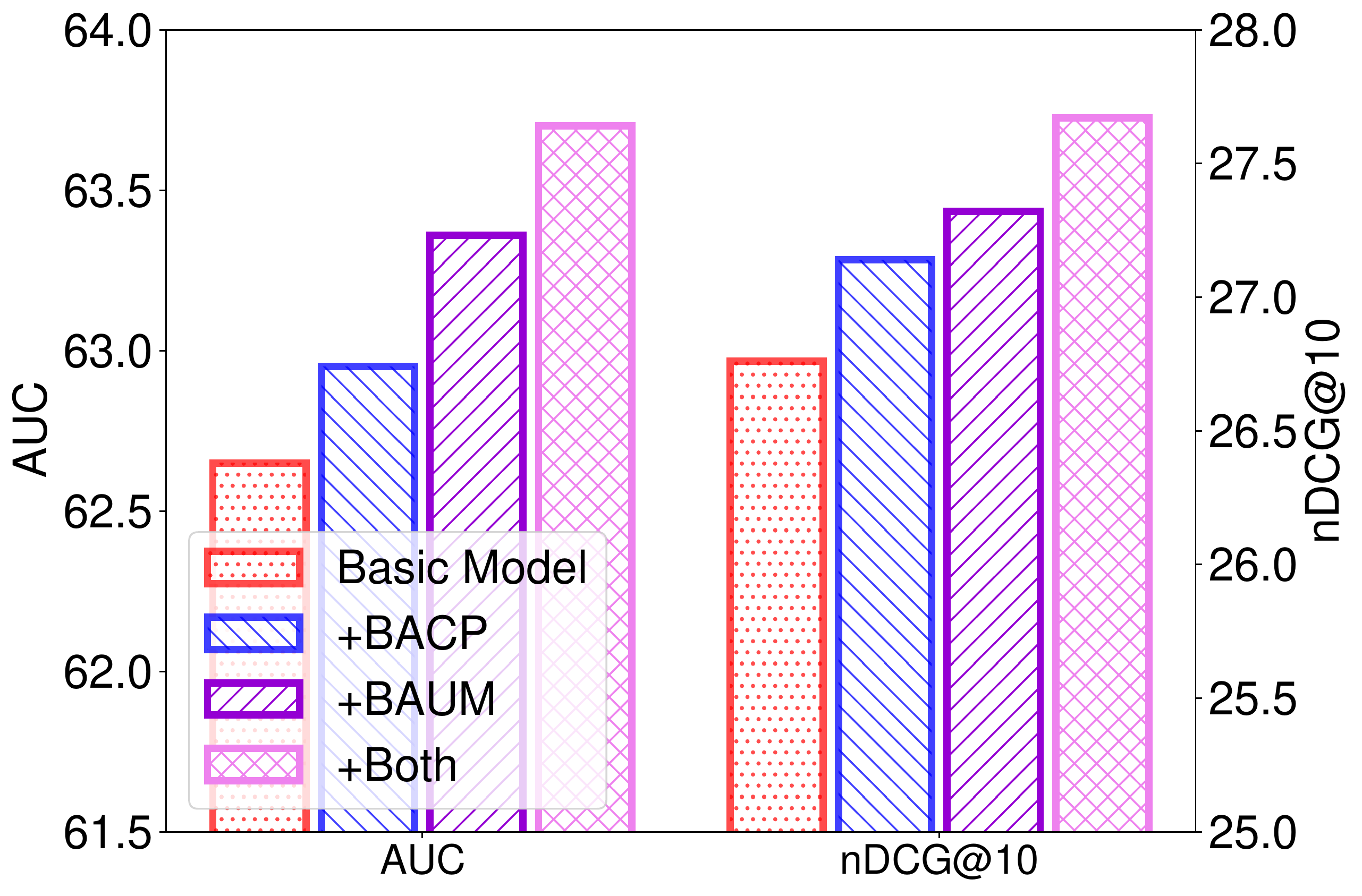}
      \caption{AUC and nDCG@10 on \textit{News}}
    \end{subfigure}
    \begin{subfigure}[b]{0.49\textwidth}
      \includegraphics[width=0.9\textwidth]{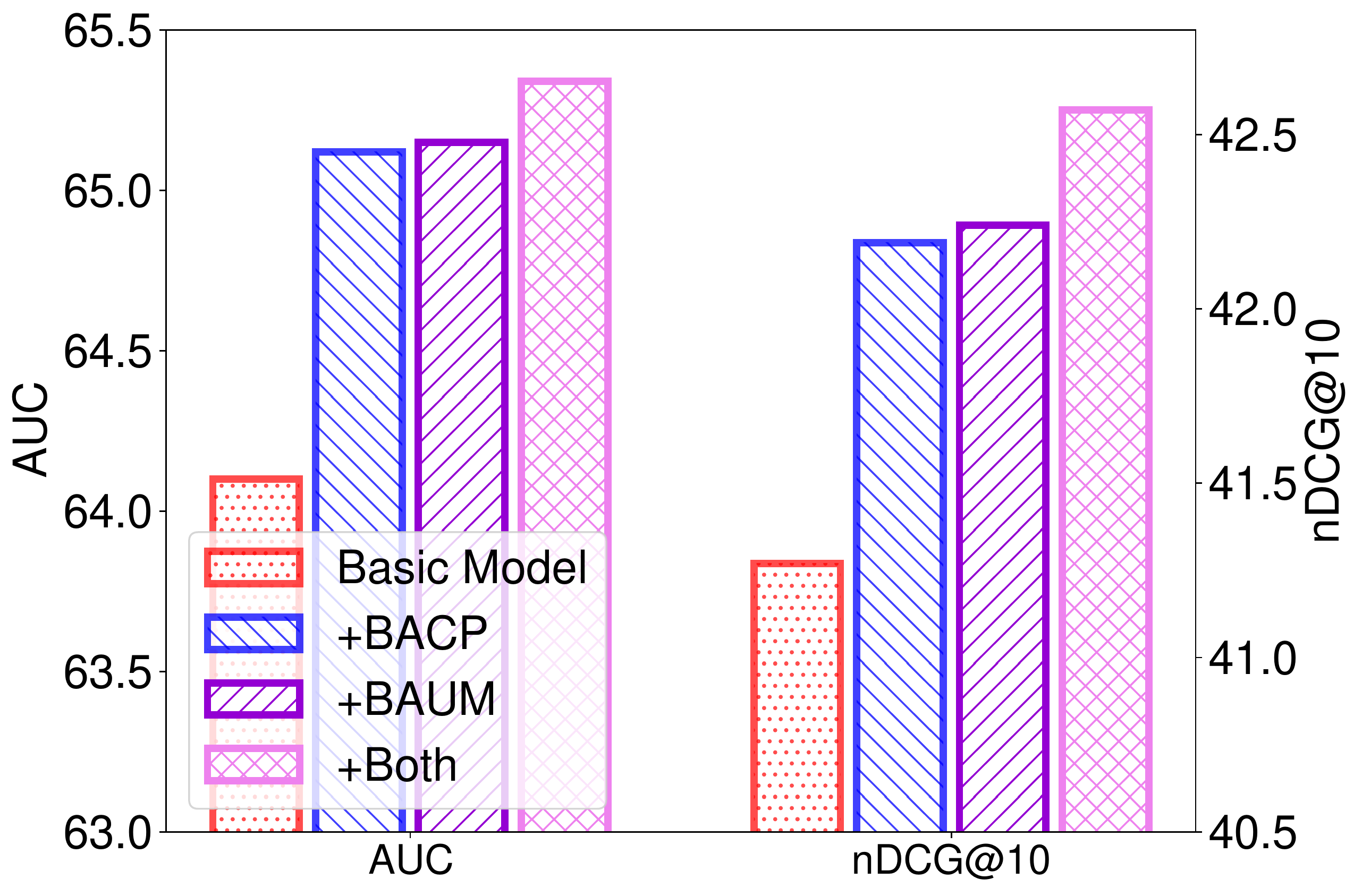}
      \caption{AUC and nDCG@10 on \textit{Feeds}}
    \end{subfigure}
    \caption{Effectiveness of BAUM and BACP.}
    \label{fig:bauecp}
\end{figure*}

\subsection{Performance Evaluation (RQ1)}

We evaluate the performance of DebiasRec by comparing it with two groups of baselines.
The first group consists of several baseline methods for news recommendation, including:
\begin{itemize}
\item \textbf{LibFM}~\cite{rendle2012factorization} a factorization machine based recommendation method.
In this paper, we use TF-IDF features of candidate news and TF-IDF features of user clicked news as inputs.
\item \textbf{DSSM}~\cite{huang2013learning} a deep structured semantic model. 
In this work, we regard candidate news as documents and user clicked news as queries.
\item \textbf{DeepFM}~\cite{guo2017deepfm} a factorization-machine based deep network, which contains a FM channel and deep channel.
\item \textbf{DFM}~\cite{lian2018towards} a multi-channel deep fusion model for personalized news recommendation. In this work, TF-IDF features of user clicked news and candidate news are treated as inputs.
\item \textbf{Wide\&Deep} ~\cite{cheng2016wide} a deep learning based recommendation method that combines a wide channel and a deep channel. TF-IDF features of candidate news and user clicked news are used as inputs.
\item \textbf{DKN}~\cite{wang2018dkn} deep knowledge-aware network for news recommendation. 
\item \textbf{NAML}~\cite{wu2019neural} a neural news recommendation methods with attentive multi-view learning.
\item \textbf{NRMS}~\cite{wu-etal-2019-neural-news-recommendation} a neural news recommendation method that uses self-attention for news and user modeling.
\end{itemize}
The second group consists of several baseline methods for click data debiasing, including:
\begin{itemize}
\item \textbf{REM}~\cite{wang2018position}: regression-based EM algorithm for unbiased learning to rank. 
We use NAML as the predictor model in REM.
\item \textbf{PAL}~\cite{guo2019pal}: position-bias aware learning framework for CTR prediction. In training stage, a position module is used for modeling position bias. In testing process, the position module is removed.
\end{itemize}
In all baseline methods, news are represented by their titles for fair comparison.
The results are shown in Table~\ref{tab:baslines}, from which we have two major findings.
First, compared with the methods (e.g., NAML and NRMS) that do not consider the bias information in news clicks, bias-aware recommendation methods (e.g., PAL and DebiasRec) perform better.
This is because the bias information in click behaviors will affect user interest modeling and model training, which is usually not beneficial for targeting user interests.
Second, our DebiasRec method consistently outperforms all baseline methods on both dataset, and the results of t-test show the improvements are significant. 
This is because our DebiasRec can alleviate the bias effect in both user modeling and click prediction, which can help learn debiased user interest representations and click prediction model.
In addition, our DebiasRec can model multiple kinds of bias (i.e. position bias and size bias in our experiments) and further consider the interactions between them.
Therefore, DebiasRec is able to better eliminate bias effect in user interest modeling and click prediction.

% The user preference and the bias effect are both considered into click prediction during the training stage.
% In the inference stage, the bias effect is removed and only the preference scores are used for recommendation.
% Finally, our DebiasRec approach outperforms existing bias-aware CTR prediction frameworks (e.g. PAL and REM). 
% This is probably because of the following two factors.
% First, our DebiasRec considers the bias effect in user modeling.
% Since a user may click the more attractive news even though he is not so interested in it, the historical click behaviors are biased.
% In the bias-aware user encoder, click behaviors are given importance scores based on both bias features and content features which indicates their informativeness towards user interest.
% Second, our DebiasRec considers different kinds of bias (i.e. position bias and size bias) and the interactions between them.
% In real-world news recommendation scenarios, user click behaviors are affected by several kinds of biases, and they are likely to be dependent on each other. 
% We verify the dependence of the position bias and the size bias in Section~\ref{sec:biasinter}.
% Our DebiasRec models the interactions between more bias factors, which can help model learn accurate bias scores in model training and judge the importance of the historical click behaviors of users more accurately in user modeling.

\begin{table*}[h]
\caption{The distribution of news in different positions and sizes.}
\begin{tabular}{c|cccccccccc}
\Xhline{1.5pt}
\hline
\multirow{2}{*}{Size} & \multicolumn{10}{c}{Position}                                                                  \\ \cline{2-11} 
                        & 1       & 2       & 3       & 4       & 5       & 6       & 7       & 8      & 9       & 10      \\ \hline
mini                   & 10,989  & 12,659  & 14,399  & 72,671  & 46,262  & 37,664  & 17,379  & 63,947 & 24,4801 & 254,513 \\
small                  & 472,250 & 283,015 & 407,429 & 298,987 & 279,691 & 29,6784 & 218,015 & 93,482 & 13,3607 & 115,277 \\
medium                 & 10,893  & 10,454  & 38,405  & 73,495  & 51,084  & 24,293  & 18,094  & 11,873 & 18,650  & 25,685  \\
large                  & 16,876  & 5,315   & 12,166  & 26,533  & 34,471  & 21,106  & 22,500  & 10,495 & 10,080  & 8,620   \\ \hline
\Xhline{1.5pt}
\end{tabular}
\label{tab:chisquare}
\end{table*}

\subsection{Effectiveness of BAUM and BACP (RQ2)}
\begin{figure*}[!t]
    \centering
    \begin{subfigure}[b]{0.49\textwidth}
      \includegraphics[width=0.9\textwidth]{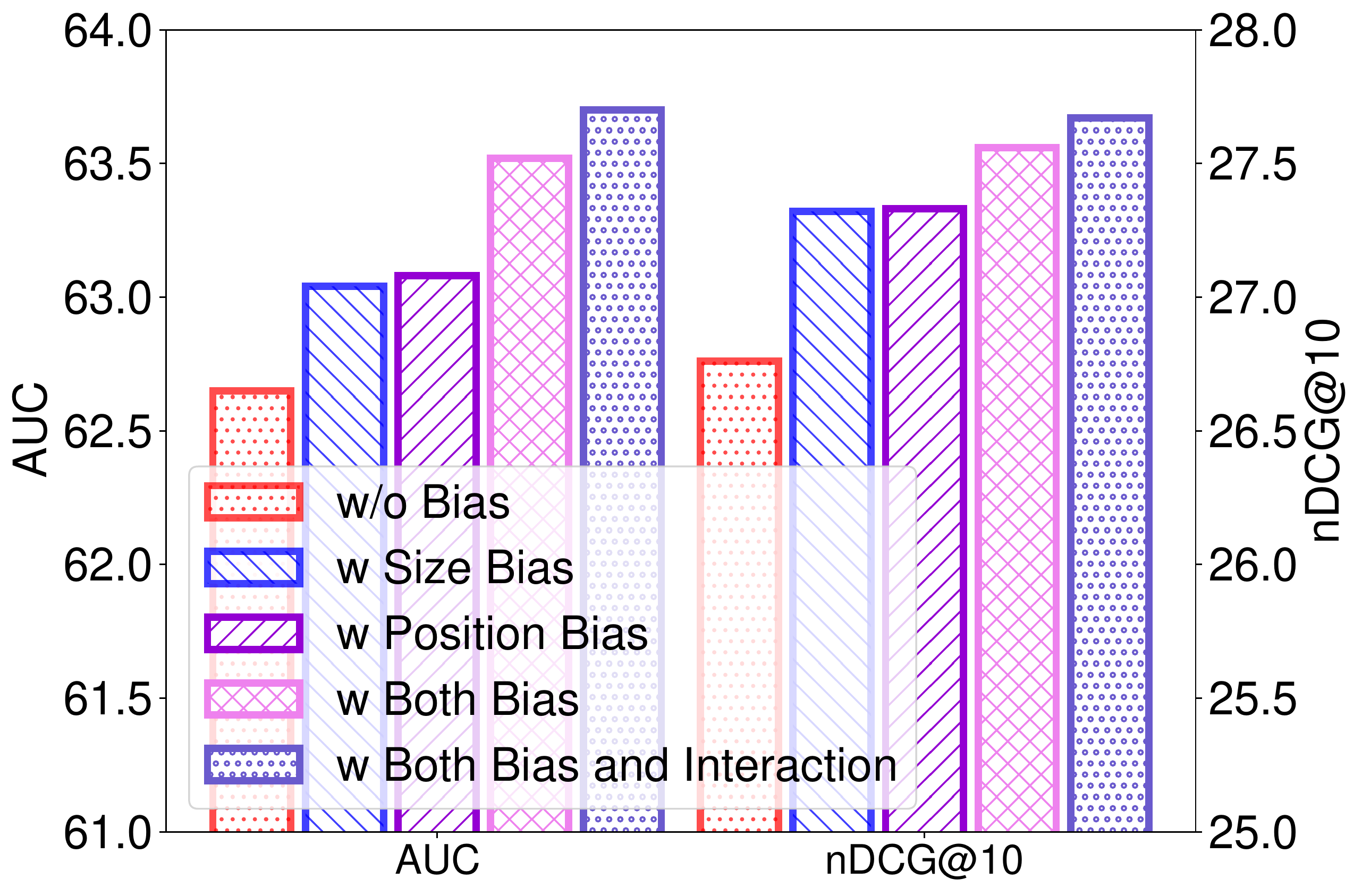}
      \caption{AUC and nDCG@10 on \textit{News}}
    \end{subfigure}
    \begin{subfigure}[b]{0.49\textwidth}
      \includegraphics[width=0.9\textwidth]{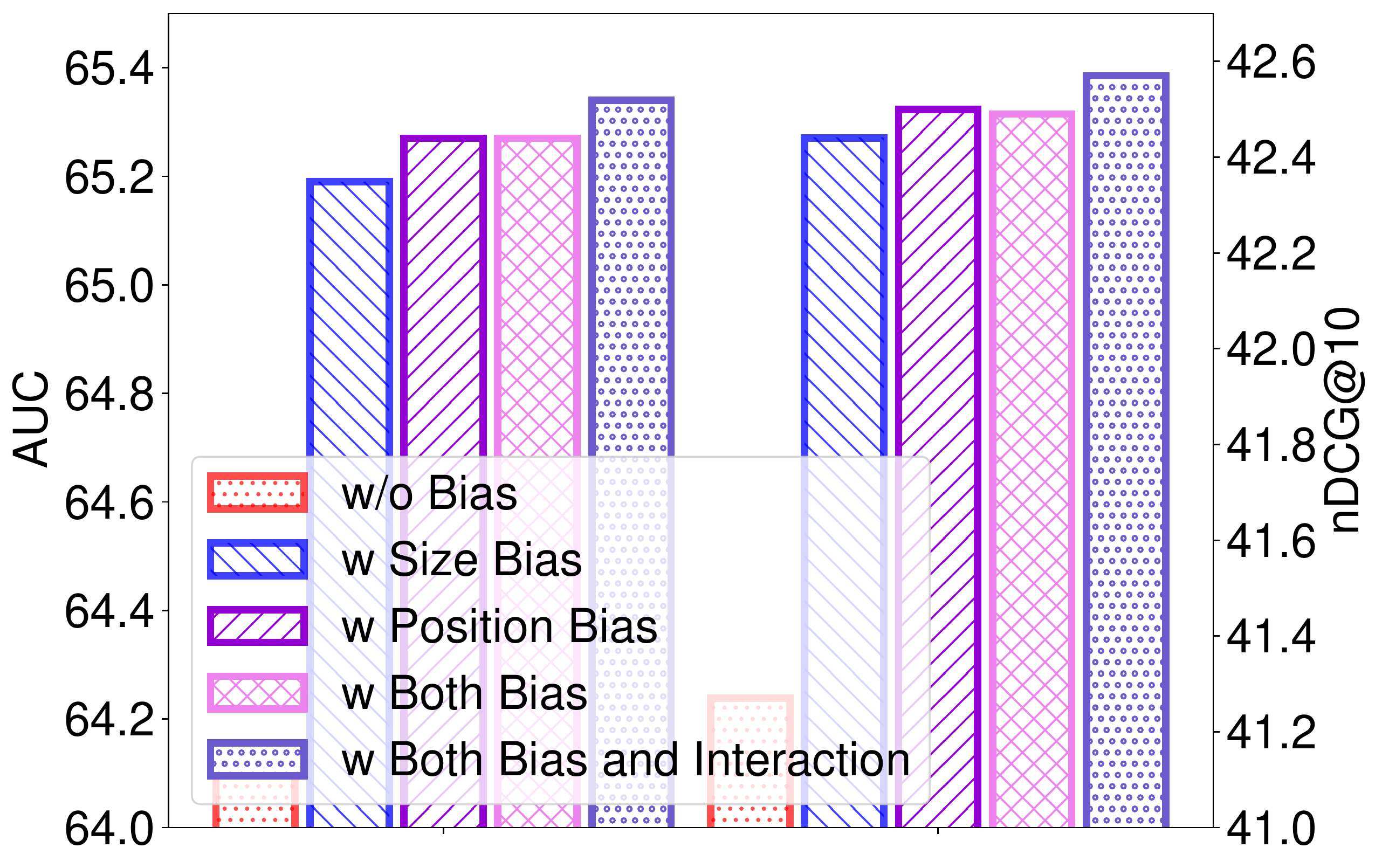}
      \caption{AUC and nDCG@10 on \textit{Feeds}}
      \label{fig:diffbias-feed}
    \end{subfigure}
    \caption{Effectiveness of modeling different kinds of biases and their interactions.}
    \label{fig:diffbias}
\end{figure*}
\begin{figure*}[!t]
  \centering
  \includegraphics[width=0.9\textwidth]{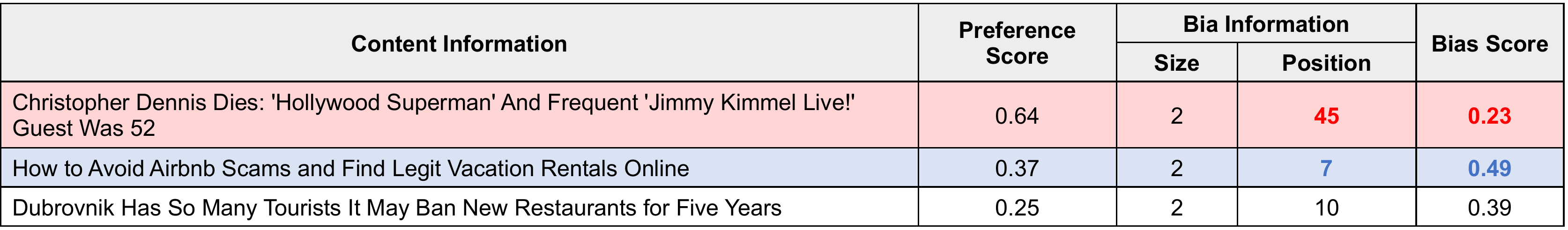}
  \caption{Visualization of preference scores and click scores in BACP.}
  \label{fig:viscore}
\end{figure*}
In this section, we further conduct several experiments to validate the effectiveness of several core components in DebiasRec, i.e., bias-aware user modeling (BAUM) and bias-aware click prediction (BACP). 
We compare the recommendation performance of DebiasRec and its variants with one of these components removed.
The results  on \textit{News} and \textit{Feeds} are shown in Figure~\ref{fig:bauecp}.
We have several observations from the results.
First, incorporating the bias-aware click prediction module can improve model performance on  both datasets.
This is because in real-world recommendation scenarios,  interest-independent biases such as position bias and size bias have strong impacts on user click behaviors.
Thus, the click labels cannot fully reflect user preference.
Our DebiasRec decomposes the click score into a bias score to model the bias effect on news click and a preference score to indicate bias-independent user interests, and we only use the preference score for news ranking.
Thus, the bias effect on click behaviors on model training can be effectively reduced.
In addition, the bias-aware user modeling can also consistently improve the performance.
This is because the historical user click behaviors for user interest model are also biased.
Our DebiasRec can further debias the historical click behaviors in user interest modeling to learn debiased user representations, which can benefit the subsequent user interest matching.
Besides, incorporating both BACP and BAUM modules can further improve the recommendation performance, which validate their effectiveness.

\subsection{Effectiveness of BRM (RQ3)}
\label{sec:biasinter}
In this section, we first demonstrate the dependency of position bias and size bias based on \textit{News} dataset. 
Then we investigate how modeling the interaction between different biases contributes to our method. 

\paragraph{Dependency between position and size}
In most scenarios, different kinds of biases are not independent to each other and they usually have joint effects on user behaviors.
To show this phenomenon, we test the dependency of position bias and size bias on the \textit{News} dataset using the Chi-square table.
Table~\ref{tab:chisquare} shows the distribution of news displayed in different positions and sizes. 
The p-value of Chi-square test is less than 0.01, which indicates the news positions and news sizes are not independent of each other on \textit{News}.\footnote{We have the same finding on the  \textit{Feeds} dataset.}

We then conduct an ablation study to verify the effectiveness of modeling multiple kinds of biases (i.e. position bias and size bias) and the interactions between them.
We compare the recommendation performance of DebiasRec on \textit{News} and \textit{Feeds} and its variants with one of these biases or their interactions removed.
We have several observations from Figure~\ref{fig:diffbias}. 
First, when the position bias is added, further incorporating the size bias can effectively improve the model performance on the \textit{News} dataset, while the  improvement is marginal on \textit{Feeds}.
After carefully studying, we find the user interface on the platform of \textit{Feeds} is static, but the user interface in \textit{News} is dynamic.
For example, the news in first position and second position in \textit{Feeds} only displayed with a fixed type ``NewsDigest''. 
The tenth position in \textit{Feeds} is assigned the displayed type ``VideoCard''.
In \textit{Feeds}, given a position $p$, the size is known.
Therefore, the results in~\ref{fig:diffbias-feed} match the real-world scenarios. 
Second, either adding position bias or size bias outperforms the method with no bias in both datasets.
It makes sense since user click behaviors are affected by both the position and the size bias in real-world recommendation scenario.
Alleviating the bias effect of either of them can help better model user interest.
Third, considering both kinds of biases can further improve the performance in \textit{News}.
This is because user click behaviors are affected by several kinds of biases in real-world recommendation scenario.
Additionally, the user interface in \textit{News} is dynamic, so considering two kinds of the bias information can help model the bias effect more precisely than considering single bias.
Forth, the bias interaction module helps in \textit{News}.
This is because the position and size are not totally independent to each other.
Compared with perceptron layer, the element-wise product in bias representation model (BRM) has better performance in exploring the interaction between different features.

% According to Figure~\ref{fig:viscore}, either adding position bias or size bias outperforms the method with no bias, which validate the effectiveness of modeling position bias and size bias.
% Comparing with only one kind of bias and with both biases, it shows considering different kinds of bias can improve performance.
% In addition, comparing methods with and without modeling bias interaction, the method with bias interaction have better performance. These results validate the effectiveness of modeling different kinds of bias and their interaction.

\begin{figure*}[!t]
  \centering
  \includegraphics[width=0.9\textwidth]{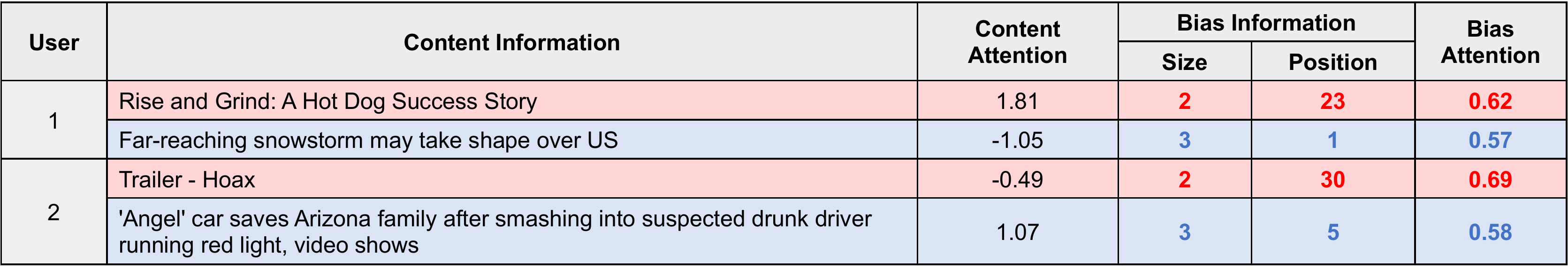}
  \caption{Visualization of attention weights in BAUM.}
  \label{fig:viwei}
\end{figure*}
\subsection{Case Study (RQ4)}

In this section, we first provide a case study on what the bias scores and preference scores will the bias-aware click predictor assign to different candidate news. 
To explore how the bias-aware predictor works, we randomly select a user who prefers clicking entertainment news, such as ``Keanu Reeves holds hands with Alexandra Grant as the Internet swoons", "Jamie Lee Curtis Opens Up About Being 20 Years Sober, Going Public With Her Addiction" and "Cailtyn Jenner, Khloe Kardashian's relationship remains strained as Caitlyn turns 70".
We randomly select three candidate news and compute their preference scores and bias scores in the bias-aware click predictor.
The first news is entertainment news about an actor, and is displayed at a lower position by the user.
The second and the third news are about vacation and travel which are irrelevant to entertainment, and they are displayed at higher positions than the first news.
The first and the second news are clicked by the user, while the third is not clicked.
The prediction results are shown in Figure~\ref{fig:viscore}.
We can observe that the bias-aware click predictor assign the highest preference score to the first news.
This is because our DebiasRec can capture user interest, and the first news is of his/her interested category.
We can also find that the first news is assigned the lowest bias score among all three news.
This is because our DebiasRec can assign bias score acording to the bias effect.
The final click score of the first and second news is higher than the third news, which show our DebiasRec can predict click behaviors more precisely after considering the bias effect.

% The first news has a high preference score and a low bias score, because it is entertainment news about an actor and is displayed at a low position and a small size.
% The second and third news have lower preference score and higher bias score, because they are about traveling and displayed at a higher position.
% These results validate the effectiveness of the bias-aware click predictor.

Then we conduct a case study to show how the bias-aware user encoder selects user clicked news based on both news contents and bias factors.
We random sample two users and two news in their historical clicked behaviors.
The results are shown in Figure~\ref{fig:viwei}.
We have several observations from the cases.
First, the bias-aware user encoder can reasonably debias the user click behaviors from the bias factors.
The first news of the first and second user are shown in small sizes and low positions, which indicates the users are interested in the news contents. 
Therefore, they are assigned high bias attention scores.
The second news of both users are shown in larger sizes and higher positions, which are assigned lower bias attention scores.
Second, the bias-aware user encoder can select more informative news to model user interest according to news content information.
For the clicked news of the first user, the first news is about food and drink which indicates the user may be interested in making food, and the second news is a regular weather forecast news which is less informative.
Therefore, our bias-aware user encoder computes a low content attention score for the first news and a high score for the second one.
For the clicked news of the second user, even though the first news is a movie name, it is hard for our model to guess user interests from two words.
Therefore, it is a assigned low content attention score.
This observation also shows our content encoder cannot understand entity information, which inspires us to perform knowledge-aware content encoder in our future work.

\section{Conclusion}

In this paper, we propose a bias-aware personalized news recommendation method named DebiasRec, which can effectively reduce the bias effect on user interest inference and model training.
The core of our approach is a bias representation model (BRM), a bias-aware user modeling model (BAUM), and a bias-aware click prediction model (BACP).
BRM is used to learn bias representation from different kinds of news biases and meanwhile model their interactions to capture their joint effect on click behaviors.
BAUM aims to model debiased user interests from clicked news articles by using their bias information to calibrate the user interest model.
BACP is used to train an debiased news recommendation model from the biased click behaviors by decomposing the click score into a preference score that indicates user interest in news content and a bias score inferred from its various bias features.
Extensive experiments on two real-world datasets show that our method can effectively improve the performance of news recommendation.


% \begin{acks}
% \end{acks}

\bibliographystyle{ACM-Reference-Format}
\bibliography{main}

% \appendix

\end{document}